\documentclass{iaus}
\usepackage{graphicx}

\title[Abundances of Bulge Giants] 
{A First Study of Giant Stars in the Galactic Bulge based on
    Crires spectra}

\author[Ryde et al.]   
{N. Ryde$^1$, %
 B. Edvardsson$^1$, B. Gustafsson$^1$, \and  H.-U. K\"aufl$^2$}

\affiliation{$^1$Department of Astronomy and Space Physics, Uppsala
University, Sweden \break email: ryde, be, bg@astro.uu.se\\[\affilskip]
$^2$ESO, Garching, Germany, email: hukaufl@eso.org}

\pubyear{2004}
\volume{xxx}  
\pagerange{119--126}
\date{?? and in revised form ??}
\jname{Proceedings Title IAU Symposium}
\editors{A.C. Editor, B.D. Editor \& C.E. Editor, eds.}
\begin{document}

\maketitle

\begin{abstract}
We present our on-going work on the determination of elemental
abundances of giants in the Galactic Bulge by means of infrared
spectroscopy. We show a preliminarily reduced spectrum and a
synthetic spectrum fit of the Bulge giant Arp 4203 recorded with the
near-infrared, high-resolution Crires spectrograph mounted on the
VLT during its science verification run in August 2006. Abundances
derived from this spectrum are discussed.

\end{abstract}


Bulges are very important building blocks of galaxies and an
exploration of them provides vital clues towards the understanding
of galaxy formation and evolution. We are involved in a project with
the goal to constrain the formation history, age, and chemical
evolution of the Milky Way Bulge and its stellar populations. We
will study the precise chemical compositions of giant stars in the
Bulge, by means of near-infrared, high-resolution spectroscopy. This
is now possible with the Phoenix and Crires spectrometers mounted on
the Gemini South and VLT telescopes, respectively. Here, we present
our first results of the Bulge giant Arp 4203 observed with Phoenix
in July and Crires in August 2006. The full analysis will be
presented later (Ryde et al. 2007, in preparation).

High-resolution, near infrared spectra of the $15530-15590$ \AA\
(the Phoenix spectrum) and the $15310-15700$ \AA\ (the new Crires
spectra) regions are recorded. The resolving powers are
$R\sim60,000$. The advantage of observing in the near-infrared is
the lower dust obscuration towards different bulge fields.
Furthermore, infrared spectra are easier to analyse than optical
spectra for metal-rich and cool stars (see for example Ryde et al.
2005). Molecular features are ubiquitous and are useful tools in the
analysis. A draw-back with an analysis in the near-infrared only, is
the difficulty in determining the stellar parameters
spectroscopically. We are investigating different methods to
determine these parameters.

We have modelled our observed spectra with synthetic spectra based
on spherical MARCS model atmospheres. The stellar parameters of the
star are $T_{\mathrm{eff}}=3900$ K, $\log g= 0.5$, $M=0.8\, \mathrm
{M_\odot}$, [Fe/H] $= -1.25$, and $\xi_{\mathrm{micro}}=2.0$
km\,s$^{-1}$ (based on the analysis of Fulbright et al. 2006). The
analysis of this rather metal-poor giant
is made relative to Arcturus. 
As an example we show one of our 4 Crires spectra of the metal-poor
bulge giant Arp 4203 in Figure 1, together with a preliminary
synthetic spectrum. Crires has four detector frames which are
recorded simultaneously and provides 4 times more wavelength
coverage and is more sensitive than the Phoenix spectrograph.

\begin{figure}
\centering \resizebox{10.9cm}{!}
 {\includegraphics{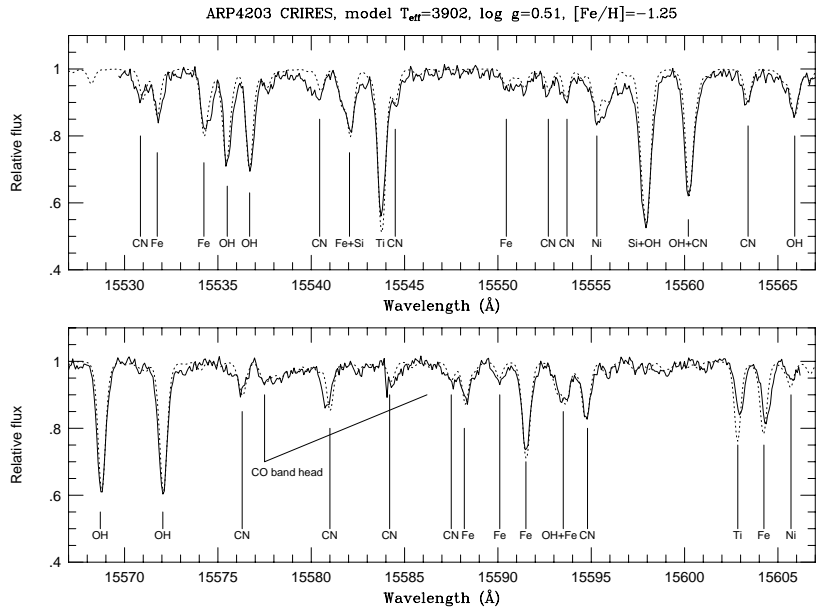}}
\caption{The third frame of our Crires spectrum of ARP4203 (full
line). Several molecular lines from CN($v=0-1$ \& $1-2$) and
OH($v=3-1$ \& $2-0$), and the band head of the second overtone of
CO($v=3-0$)
are modelled, as well as lines from Fe, Ni, Si, and Ti (dashed line).
} \label{fig:wave}
\end{figure}
%

%

\begin{table}\def~{\hphantom{0}}
{\scriptsize \begin{center}
  \caption{Preliminary stellar abundances derived for Arp 4203. Uncertainties are of the order of 0.1 dex. The standard solar abundances are taken from Grevesse \&
Sauval 1998, apart from the C, N, \& O abundances which are taken
from Asplund et al. 2003. }
  \label{tab2}
  \begin{tabular}{lccccccc}\hline
Element &\multicolumn{2}{c}{ARP 4203 (this study)}  &  \multicolumn{2}{c}{ARP 4203 (Fulbright et al. 2006)} & \multicolumn{2}{c}{Arcturus ($\alpha$ Boo)} & Solar abundnace\\
 & $\log\varepsilon$ & $\Delta\log\varepsilon^\ddag$ & $\log\varepsilon$ & $\Delta \log\varepsilon^\ddag$ & $\log\varepsilon$ & $\Delta \log\varepsilon^\ddag$ &$\log\varepsilon$\\
\hline
C & $6.65$ & $-0.51$ &  $-$ & $-$ & $8.02$ &  $+0.11$ & $8.41$  \\
N & $7.75$ & $+1.20$ &  $-$ & $-$ & $7.59$ &$+0.29$ & $7.80$\\
O & $7.75$ & $-0.06$ & $7.69$ & $-0.12$  & $8.67$ &$+0.31$ & $8.66$\\
C+N+O & $8.07$ & $+0.15$ &  $-$ & $-$ & $8.79$  & $+0.27$& $8.89$\\
Si & $6.70$  & $\pm0.00$ & $6.87$ & $+0.17$ & $7.25$ & $$ & $7.55$ \\
Ca & $5.51$ & $\pm0.00$ & $5.45$ & $-0.06$ & $6.06$ & $$ & $6.36$  \\
Ti &  $4.17$ & $\pm0.00$ & $4.05$ & $-0.12$ & $4.72$&  & $5.02$  \\
Fe &  $6.30$ & $+0.05$ & $6.22$ & $-0.03$ & $6.96$ & & $7.50$  \\
Ni &  $5.17$ & $+0.17$ & $-$ & $-$ & $5.73$ & & $6.25$ \\

\hline
  \end{tabular}
 \end{center}
{\scriptsize
 $^\ddag$ The abundance compared to the scaled solar one, accounting for the $\alpha$-element enhancement, $\mathrm{[\alpha/Fe]}=0.4$.}}
\end{table}
%


In Table~\ref{tab2} we provide our preliminary derived elemental
abundances for Arp 4203. The giant is depleted in C and enriched in
N, whereas the abundances of O has not changed much, which are signs
that matter exposed to the CN cycle (which conserves the sum of C
and N nuclei) has been dredged up to the stellar surface. The sum of
the abundances of C, N, and O is close to that expected from a
non-processed star. We find the abundances of the $\alpha$-elements
Si, Ca, and Ti to be $\mathrm{[\alpha/Fe]}=0.4$, at the expected
levels for an $\alpha$-enhanced metal-poor star. The nickel
abundance is slightly higher than the overall metallicity. Our
measurements lie close to those of Fulbright et al. (2006) for the
elements we have in common. A full analysis and discussion will be
presented in Ryde et al. 2007 (in preparation).


\end{document}